\documentclass[compsoc,journal]{IEEEtran}
\usepackage[T1]{fontenc}
\usepackage{amsmath,amsfonts,amssymb}
\usepackage{algorithm}
\usepackage{algpseudocode}
\usepackage{array}
\usepackage[caption=false,font=footnotesize]{subfig}
\usepackage{textcomp}
\usepackage{stfloats}
\usepackage{url}
\usepackage{verbatim}
\usepackage{graphicx}
\usepackage{cite}
\usepackage{booktabs}
\usepackage{multirow}
\usepackage{siunitx}
\usepackage{tabularx}
\usepackage{xcolor}
\usepackage{xspace}
\usepackage{tikz}
\usetikzlibrary{positioning,arrows.meta,fit,backgrounds}
\usepackage[hidelinks]{hyperref}
\graphicspath{{figure/}}
\newcolumntype{L}[1]{>{\raggedright\arraybackslash}p{#1}}

\newcommand{\sysname}{WiSI\xspace}
\newcommand{\Ideal}{\textsc{Ideal}\xspace}
\newcommand{\Radio}{\textsc{Radio-map}\xspace}
\newcommand{\Unif}{\textsc{Uniform}\xspace}
\newcommand{\UnifHold}{\textsc{Uniform-hold}\xspace}
\newcommand{\Mixed}{\textsc{Mixed}\xspace}
\newcommand{\Trace}{\textsc{Trace}\xspace}

\algrenewcommand\algorithmicensure{\textbf{Output:}}

\begin{document}
\bstctlcite{IEEEtranBSTCTL_NoDash}

\title{WiSI: Wireless-in-the-Loop Self-Improvement}

\author{Zongxi Liu,
    and~Hongyang Du$^*$
    \thanks{Zongxi Liu and Hongyang Du are with the Department of Electrical and Computer Engineering, The University of Hong Kong, Hong Kong SAR, China (e-mail: lzxixi@connect.hku.hk; duhy@hku.hk).}}

\maketitle

\vspace{-1cm}
\begin{abstract}
    The growing autonomy of Artificial Intelligence (AI) agents has enabled self-improvement, where agents iteratively refine their programs through execution feedback and performance evaluation. As these agents increasingly operate in physical environments and rely on wireless connectivity, communication conditions become an integral part of their improvement process. However, existing approaches largely assume ideal communication, overlooking how wireless conditions shape both the feedback that guides program modifications and the evaluation outcomes that determine which modifications survive. This challenge is further complicated by the coupling between agent behavior and wireless connectivity, as agents influence their own communication conditions through mobility and transmission decisions. To address this problem, we propose WiSI, a wireless-in-the-loop self-improvement framework that integrates spatiotemporal wireless channels, edge-assisted perception, and Large Language Model (LLM)-driven program improvement. By incorporating wireless execution into the improvement loop, WiSI reveals how wireless execution experience shapes the generation and retention of program structures, and investigates the wireless conditions needed for robust improvement. Experiments show that agents improved under ideal links achieve only 40.0\% task success under congestion, compared with 77.2\% when improved under realistic wireless conditions. More importantly, wireless-aware feedback and selection jointly enable agents to discover and retain communication-efficient program structures that outperform parameter tuning and remain robust even with inaccurate run-time channel observations. We further find that moderate-load or mixed-network evaluation maintains strong performance across deployment conditions, while simplified models preserving link-rate statistics achieve comparable results without detailed spatial channel information.
    The code is available at \url{https://github.com/NICE-HKU/WiSI}.
\end{abstract}

\begin{IEEEkeywords}
    Self-improving agents, edge intelligence, Mobile AI agents, wireless networks.
\end{IEEEkeywords}

\IEEEpeerreviewmaketitle

\section{Introduction}\label{sec:intro}

\IEEEPARstart{A}{utonomous} artificial intelligence (AI) agents are evolving from systems with predefined execution logic toward systems capable of improving their behavior through experience. Such agents may rely on learned models, executable programs, or their combination to coordinate perception, reasoning, and actions. Recent advances in large language models (LLMs) have enabled program-level self-improvement, where an improvement operator iteratively rewrites an agent's execution program based on its task experiences, evaluates candidate modifications, and retains successful versions~\cite{zelikman2024stop,hu2025adas,zhang2025dgm,wang2025hgm}.  This capability is especially relevant to embodied intelligence, where agents operate in diverse physical environments and face tasks that cannot be fully anticipated during development~\cite{wang2023voyager,yang2025embodiedbench}. Applications such as autonomous inspection and warehouse robotics require mobile agents to coordinate physical actions with computationally intensive perception and reasoning. Since these capabilities often exceed onboard resources, wireless connectivity to edge and remote computing infrastructure becomes a key enabler ~\cite{zhou2019edge,kehoe2015cloudrobotics,mao2017mec,xu2026collaboration}. Incorporating program-level self-improvement into such wireless-connected agents could reduce manual adaptation across tasks and deployment conditions.

\begin{figure}[t]
    \centering
    \includegraphics[width=\columnwidth]{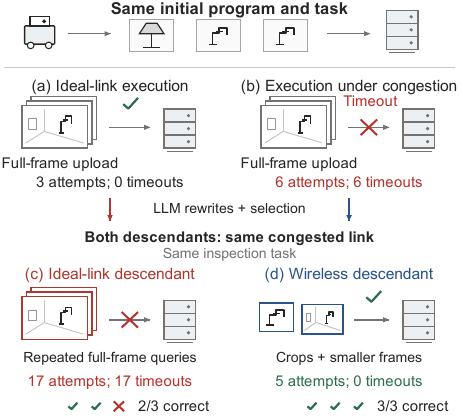}
    \caption{Motivation for wireless-in-the-loop self-improvement.
        (a), (b) The same initial agent program executes an inspection task under ideal and congested wireless conditions, yielding different execution feedback.
        (c), (d) The resulting improved programs exhibit different communication behaviors. Under a common congested network, the ideal-link-improved program relies on repeated full-frame uploads, whereas the wireless-in-the-loop-improved program adopts compact uploads and achieves higher task success.}
    \label{fig:motivation}
\end{figure}

Despite this promise, the role of wireless communication in agent self-improvement remains largely unexplored. Existing approaches usually improve agents under idealized or unspecified communication conditions, even though wireless connectivity directly affects both execution feedback and evaluation outcomes. Fig.~\ref{fig:motivation} illustrates the core issue through an inspection task. The same initial program is first executed under an ideal link and under a congested link. Under ideal communication, full-frame uploads complete successfully while under congestion, the same behavior repeatedly times out. These different execution experiences guide the generation and selection of different program modifications, leading to distinct improvement trajectories. When the two descendants are replayed on the same inspection task under a common congested network, the ideal-link descendant continues to rely on repeated full-frame queries, producing 17 upload attempts, all of which time out, and answers only 2 of the 3 inspection items correctly. By contrast, the wireless descendant adopts crops and smaller frames, completes all 5 uploads without timeout, and answers all 3 items correctly. This example illustrates the central question: how do wireless execution experiences influence the program structures discovered and retained through self-improvement?

This question cannot be answered by viewing wireless communication as a simple source of delay or noise during task execution. In self-improving agents, the execution environment influences the experiences from which new candidate programs are proposed, and it also influences the scores used to decide which candidates survive. These two roles are conceptually distinct. Even if candidate selection is conducted under a realistic network, useful behaviors may still fail to emerge if the preceding execution logs were collected under ideal communication and therefore do not expose relevant failures. The problem is further complicated by the fact that an embodied agent affects its own communication conditions through navigation, upload timing, and payload choices, so the evolving program and the wireless environment are coupled~\cite{ghaffarkhah2011motion}. Moreover, evaluation and deployment conditions may differ, and constructing detailed wireless models for every target site can be costly~\cite{sorour2015joint,romero2022radiomap}. Studying how wireless conditions steer program evolution therefore requires separating their effects on feedback generation and candidate selection while preserving their coupling with agent behavior.

To address this problem, we propose \emph{WiSI}, a wireless-in-the-loop self-improvement framework that explicitly incorporates wireless communication into iterative program improvement. WiSI distinguishes two roles of the wireless environment in the improvement loop. Execution feedback exposes failures and outcomes that guide program modifications, while evaluation scores determine which modifications are retained and become the starting point for later generations. At the same time, WiSI couples agent execution with spatiotemporal wireless conditions, so changes in navigation and communication behavior alter the communication experiences encountered during later executions. The illustrative case in Fig.~\ref{fig:motivation} previews this intuition. We systematically investigate how these two mechanisms affect program evolution through controlled experiments and multiple improvement lineages.

We conduct extensive experiments to investigate these effects across different wireless conditions and deployment environments. We first examine how network mismatch affects fixed strategies and candidate selection, and then trace its influence through successive generations of program improvement. To identify the underlying mechanisms, we independently vary the feedback and selection environments and compare the resulting programs with those obtained through conventional parameter optimization. The results show that programs improved under realistic wireless conditions achieve 77.2\% task success under congestion, compared with only 40.0\% for programs improved over an ideal link. Our feedback-selection experiments further show that realistic execution feedback changes which program structures are generated even when candidate selection already uses realistic wireless conditions. The resulting communication-efficient programs outperform parameter tuning within predefined strategy templates and remain robust under inaccurate run-time channel observations. We also identify evaluation environments and wireless models that support robust improvement across deployment conditions.
Our main contributions are summarized as follows:
\begin{itemize}
    \item We investigate how wireless conditions shape the self-improvement of embodied AI agents beyond their immediate effects on task execution. We identify two coupled mechanisms: execution feedback influences which program structures are generated, while wireless evaluation determines which modifications are retained and inherited. We formulate this process as a constrained optimization problem that maximizes deployment utility under potentially mismatched improvement and deployment conditions.

    \item We develop \sysname, a wireless-in-the-loop self-improvement framework that integrates behavior-dependent wireless execution with LLM-driven program rewriting. By incorporating wireless execution into program generation and selection while preserving program inheritance across generations, \sysname enables systematic investigation of how wireless conditions influence program evolution.

    \item We demonstrate that wireless execution feedback shapes candidate program structures from the first generation, beyond its influence on candidate selection. Under congested deployment, \sysname achieves 77.2\% task success compared with 40.0\% for ideal-link improvement, with realistic feedback providing a further 23.1-percentage-point gain even when wireless-aware selection is fixed. The evolved programs exhibit communication-efficient behaviors that outperform parameter optimization within predefined strategy templates. We further identify evaluation conditions supporting robustness across wireless loads and show that simplified wireless models can guide effective program improvement.
\end{itemize}

The remainder of this paper is organized as follows. Section~\ref{sec:related} reviews related work. Section~\ref{sec:problem} presents the system model and formulates agent self-improvement in wireless environments. Section~\ref{sec:system} analyzes wireless-dependent program evolution and presents the \sysname framework and its implementation. Section~\ref{sec:results} evaluates the effectiveness, communication efficiency, and robustness of \sysname. Section~\ref{sec:conclusion} concludes the paper.

\section{Related Work}\label{sec:related}

\subsection{Self-Improving AI Agents}
Research on self-improving AI agents has progressed from theoretical self-modification toward empirical program improvement driven by large language models (LLMs). The G\"odel machine~\cite{schmidhuber2007godel} introduced provable self-modification, while STOP~\cite{zelikman2024stop}, ADAS~\cite{hu2025adas}, and the Darwin G\"odel Machine~\cite{zhang2025dgm} enable agents to improve their scaffolds, program designs, and code repositories through iterative evaluation. More recent studies explore improvement potential beyond immediate performance~\cite{wang2025hgm} and efficient feedback through experience replay~\cite{zheng2026dreamrsi}. In embodied environments, research has similarly advanced from skill acquisition and execution reflection~\cite{wang2023voyager,shinn2023reflexion} toward program-level self-improvement~\cite{adarsi2026,robofoundry2026,shaper2026}. Related studies investigate feedback noise~\cite{rsebench2026}, degradation from self-generated data~\cite{shumailov2024collapse}, and reward misalignment~\cite{skalse2022rewardhacking}, highlighting the importance of reliable evaluation. However, research has paid limited attention to how communication environments shape the execution feedback and candidate selection that drive program improvement, particularly when agents interact with wireless infrastructure during execution.

\subsection{Wireless-Connected Embodied Intelligence}
Wireless connectivity enables resource-constrained embodied agents to access remote perception and computation through mobile edge computing and cloud robotics~\cite{mao2017mec,zhou2019edge,kehoe2015cloudrobotics,eshratifar2021jointdnn}. Research has progressed toward communication-efficient intelligence through computation partitioning~\cite{kang2017neurosurgeon,li2024distributed}, efficient model inference~\cite{leviathan2023speculative,zhang2024dvfo}, task-oriented communication~\cite{gunduz2023beyondbits}, and importance-aware transmission~\cite{liu2021importance}. Communication-aware robotics further integrates wireless conditions into physical decision-making through mobility optimization and radio-map-assisted planning~\cite{muralidharan2021commaware,romero2022radiomap,ghaffarkhah2011motion}. These studies establish the importance of jointly considering communication, computation, and physical actions, but primarily optimize execution within predefined policies or decision frameworks~\cite{li2020edgeai}. Meanwhile, embodied evaluation benchmarks such as EQA~\cite{das2018eqa}, OpenEQA~\cite{majumdar2024openeqa}, and EmbodiedBench~\cite{yang2025embodiedbench}, supported by AI2-THOR~\cite{kolve2017ai2thor} and ProcTHOR~\cite{deitke2022procthor}, generally abstract away wireless communication. Consequently, how wireless conditions influence the iterative improvement of embodied agent programs remains insufficiently understood.

\subsection{Robust Evaluation and Optimization}
Robust evaluation and optimization address performance uncertainty arising from limited observations and changing environments. Classical optimization under uncertainty~\cite{jin2005uncertain} and ranking-and-selection methods~\cite{kim2006selecting,karnin2013almost} improve candidate selection under noisy evaluations, while domain randomization~\cite{tobin2017domain} exposes learning systems to diverse operating conditions. Although these approaches improve evaluation reliability and generalization, additional sampling cannot eliminate systematic mismatch between evaluation and deployment environments. More importantly, reliable selection among existing candidates does not ensure that useful alternatives will be generated. This limitation is particularly relevant to LLM-driven program improvement, where execution feedback influences candidate generation and retained modifications shape subsequent generations. Unlike parameter optimization within predefined strategy templates, program-level improvement can explore new communication and decision-making structures. Understanding its robustness therefore requires examining how wireless conditions jointly influence the generation, selection, and inheritance of agent programs.
\section{System Model and Problem Formulation}
\label{sec:problem}

We consider embodied agents that depend on wireless communication during task execution and formulate the objective of improving their programs under limited evaluation resources. The model captures how agent behavior affects the communication conditions experienced during execution and how evaluation conditions influence the quality of the resulting programs.

\subsection{Wireless-Connected Embodied Agents}

\begin{figure}[t]
    \centering
    \includegraphics[width=0.48\textwidth]{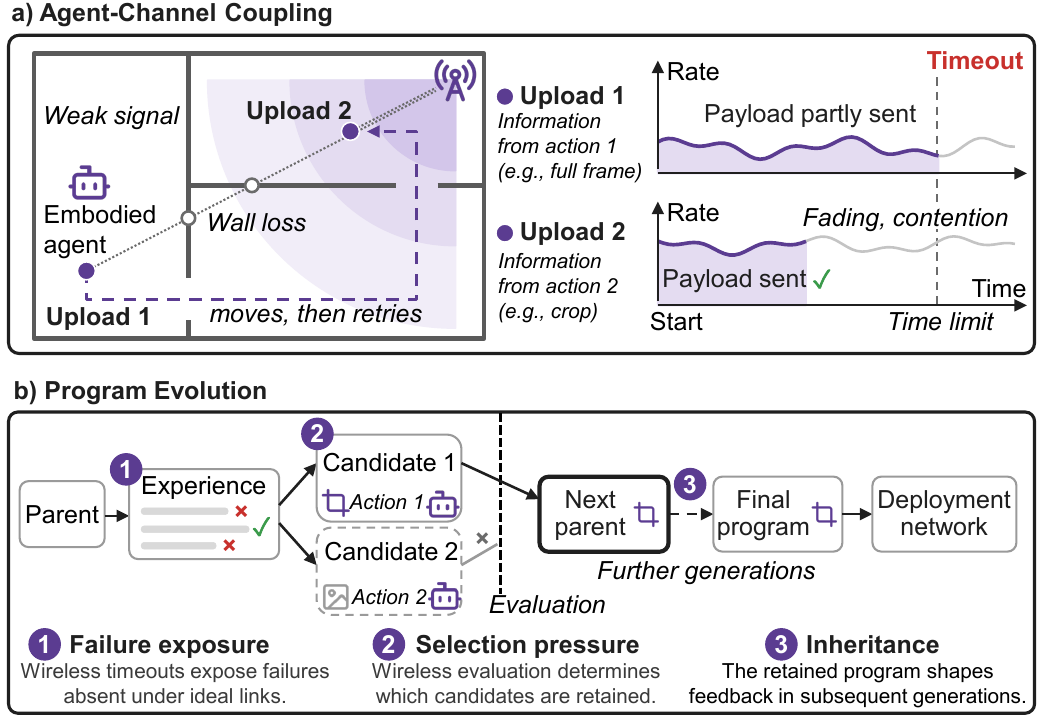}
    \caption{Wireless execution and its implications for program evolution. (a) Agent mobility and transmission decisions affect communication outcomes under time-varying wireless conditions. (b) Execution feedback influences candidate program generation, wireless evaluation determines program retention, and inherited programs shape subsequent improvement generations.}
    \label{fig:mechanism}
\end{figure}

As illustrated in Fig.~\ref{fig:mechanism}(a), we consider an embodied agent operating in a physical environment and performing tasks $\tau\sim\mathcal{T}$, where $\mathcal{T}$ denotes the task distribution. Each task specifies an objective, an environment configuration, and a completion deadline $T_\tau$. The agent is represented by an executable program $a\in\Pi$, where $\Pi$ is the space of admissible programs. Its execution policy $\pi_a$ maps observations and internal states to actions involving physical interaction, local computation, and communication with external computing resources. Observations may include sensory information, task progress, and wireless link measurements. Let $\xi$ denote a realization of the wireless environment over a task, including propagation effects, temporal fading, and network contention. The task reward is
\begin{equation}
    R(a,\tau,\xi)=
    \mathbb{1}\{\text{task completed correctly before }T_\tau\},
    \label{eq:reward}
\end{equation}
where $\mathbb{1}\{\cdot\}$ equals one when its condition holds and zero otherwise. The formulation also accommodates general task-dependent utility functions.

Wireless communication couples the agent's actions with its access to external intelligence. Let $\mathbf{x}_t$ be the agent's position at time $t$, and let $r(\mathbf{x}_t,t;\xi)$ denote the achievable uplink rate at that position and time under wireless realization $\xi$. The execution program determines the agent's movement and transmission decisions, thereby determining which portions of the wireless environment it encounters. In the indoor embodied scenario considered here, the agent accesses edge-hosted perception models through a wireless connection. Its action space is
\begin{equation}
    \begin{aligned}
        \mathcal{U}=\{ &
        \textsc{Move},\textsc{Rotate},\textsc{LocalPerceive}, \\
                       & \textsc{EdgeQuery}(b,\theta),
        \textsc{Wait},\textsc{Answer}\},
    \end{aligned}
    \label{eq:actions}
\end{equation}
where $\textsc{EdgeQuery}(b,\theta)$ requests edge perception by transmitting a payload of $b$ bits subject to an upload timeout $\theta$. The remaining actions support navigation, local perception, waiting, and task completion.

We instantiate the wireless environment using a spatiotemporal channel model~\cite{cost231}. The received signal-to-noise ratio (SNR) in decibels is
\begin{equation}
    \gamma_{\mathrm{dB}}(\mathbf{x}_t,t)
    =P_{\mathrm{tx}}-\mathrm{PL}(\mathbf{x}_t)
    -S(\mathbf{x}_t)+F_t-N_0,
    \label{eq:snr}
\end{equation}
where $P_{\mathrm{tx}}$ and $N_0$ are the transmit and receiver noise powers in dBm, respectively. The terms $\mathrm{PL}(\mathbf{x}_t)$, $S(\mathbf{x}_t)$, and $F_t$ represent path loss, spatially correlated shadowing loss, and time-varying fading gain in dB. The path loss follows a multi-wall propagation model~\cite{cost231}, given by
\begin{equation}
    \mathrm{PL}(\mathbf{x})
    =\mathrm{PL}_0
    +10\nu\log_{10}\frac{d(\mathbf{x})}{d_0}
    +\sum_{w\in\mathcal{W}(\mathbf{x})}L_w,
    \label{eq:pathloss}
\end{equation}
where $\mathrm{PL}_0$ denotes reference path loss at distance $d_0$, $\nu$ is the path-loss exponent, and $d(\mathbf{x})$ is the distance from position $\mathbf{x}$ to the access point. The set $\mathcal{W}(\mathbf{x})$ contains obstacles along the propagation path, each with penetration loss $L_w$. The resulting uplink rate is
\begin{equation}
    r(\mathbf{x}_t,t;\xi)=
    \min\left\{
    r_{\max},
    \rho\eta W\log_2
    \left(1+\gamma(\mathbf{x}_t,t)\right)
    \right\}
    \mathbb{1}\{z_t=\mathsf{G}\},
    \label{eq:rate}
\end{equation}
where $\gamma(\mathbf{x}_t,t)=10^{\gamma_{\mathrm{dB}}(\mathbf{x}_t,t)/10}$ is the SNR in linear scale, $W$ is the bandwidth, $\eta$ is the implementation efficiency, and $r_{\max}$ is the maximum uplink rate. The airtime share $\rho\in(0,1]$ represents network contention. The variable $z_t$ denotes an optional Gilbert--Elliott outage state~\cite{gilbert1960burst,elliott1963burst}, with $\mathsf{G}$ indicating link availability.

For an upload initiated at time $t_0$ while the agent remains at position $\mathbf{x}_{t_0}$, the payload is delivered within timeout $\theta$ if
\begin{equation}
    \int_{t_0}^{t_0+\theta}
    r(\mathbf{x}_{t_0},u;\xi)\,{\rm d}u\ge b.
    \label{eq:completion}
\end{equation}
Transmission outcomes and subsequent edge-inference and response-delivery delays contribute to task completion time. Different programs may therefore experience different communication outcomes under the same underlying wireless realization because they communicate at different positions and times.

\subsection{Self-Improvement in Wireless Environments}
\label{sec:rsi-formal}

Agent programs may be evaluated under conditions that differ from those encountered during deployment. Let $E$ denote an evaluation environment that specifies a distribution of wireless realizations, and let $D$ denote the corresponding deployment environment. The expected utility of program $a$ under $E$ is
\begin{equation}
    U_E(a)=
    \mathbb{E}_{\tau\sim\mathcal{T},\,\xi\sim E}
    [R(a,\tau,\xi)],
    \label{eq:utility}
\end{equation}
with deployment utility $U_D(a)$ defined analogously. For a fixed candidate set $\mathcal{C}\subseteq\Pi$, selection under $E$ chooses $\hat a=\arg\max_{a\in\mathcal{C}}\widehat U_E(a)$, where $\widehat U_E(a)$ denotes the empirical utility estimated from $N$ evaluation episodes. The performance of the selected program under $D$ can differ from that of the best candidate for deployment, giving rise to selection regret
\begin{equation}
    \mathrm{Reg}
    =\max_{a\in\mathcal{C}}U_D(a)-U_D(\hat a).
    \label{eq:regret}
\end{equation}
Increasing the evaluation budget can reduce sampling uncertainty but cannot eliminate systematic differences between evaluation and deployment conditions.

Program-level self-improvement extends candidate selection by allowing the candidate programs themselves to change through execution experience. Starting from a seed program $a_0\in\Pi$, an automated improvement procedure evaluates and modifies programs over successive iterations and produces a final program $p_G\in\Pi$ after $G$ generations. Different procedures may select from predefined candidate pools, optimize parameters within fixed strategy templates, or generate new executable programs. Their common objective is to obtain a program that performs well under the target deployment environment.

Let $\mathcal{S}\in\mathfrak{S}$ denote an admissible program improvement procedure, where $\mathfrak{S}$ is the class of procedures under consideration. Let $\mathcal{E}_{\mathrm{imp}}\in\mathfrak{E}$ specify the conditions available during improvement, where $\mathfrak{E}$ denotes the admissible choices of improvement conditions. The procedure $\mathcal{S}$ produces $p_G$ from $a_0$ using these conditions. Its operation may be stochastic, with $\omega$ denoting the randomness in program generation and evaluation. The common design objective can be expressed as
\begin{equation}
    \begin{aligned}
        \max_{\mathcal{S}\in\mathfrak{S},\,\mathcal{E}_{\mathrm{imp}}\in\mathfrak{E}}
        \quad & \mathbb{E}_{\omega}\!\left[U_D(p_G)\right]                        \\
        \mathrm{s.t.}\quad
              & p_G=\mathcal{S}(a_0;\mathcal{E}_{\mathrm{imp}},\omega)\in\Pi,     \\
              & \operatorname{Cost}(\mathcal{S},\mathcal{E}_{\mathrm{imp}})\le B,
    \end{aligned}
    \label{eq:self-improvement-optimization}
\end{equation}
where $B$ is the available improvement budget and $\operatorname{Cost}(\cdot)$ accounts for the resources used in generating and evaluating programs. The output $p_G$ is determined by the chosen procedure and its execution history. This formulation states the common objective of program improvement without prescribing how execution feedback or candidate evaluation must be organized.

The improvement conditions affect both the evidence available to a procedure and the performance estimates used to guide its search. Conventional software-based improvement may abstract away communication, even when the resulting agent later depends on wireless connectivity. In wireless-connected embodied agents, the experienced channel also depends on the program's actions. These dependencies motivate a method that explicitly incorporates wireless execution into program improvement.

\section{The \sysname Framework and Implementation}
\label{sec:system}

\begin{figure}[t]
    \centering
    \includegraphics[width=0.48\textwidth]{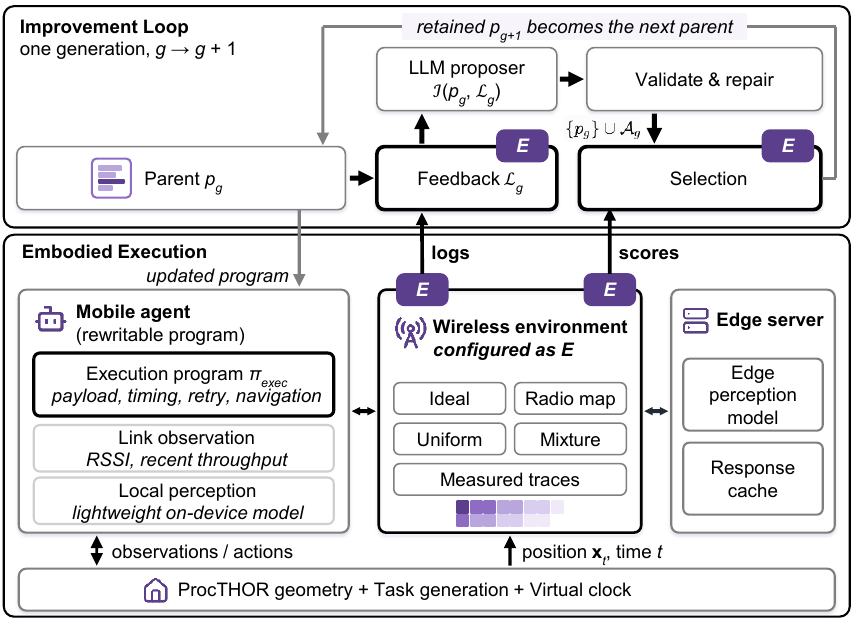}
    \caption{Architecture of the \sysname framework.
        Execution under wireless environment $E$ supplies feedback for program generation and performance scores for candidate selection.
        The retained program becomes the next parent, and its actions determine the wireless conditions encountered during execution.}
    \label{fig:arch}
\end{figure}

We develop \sysname to incorporate wireless execution conditions into iterative program improvement. The framework preserves the coupling between agent behavior and communication outcomes while making the roles of execution feedback and candidate selection explicit. As illustrated in Fig.~\ref{fig:arch}, it connects wireless evaluation with program generation and retention across successive improvement generations.

\subsection{Wireless-in-the-Loop Self-Improvement}

\sysname improves agent programs through repeated execution and evaluation under a wireless environment $E$. Execution feedback guides the generation of candidate programs, while their evaluated performance determines which program is retained for the next generation. At generation $g\in\{0,\ldots,G-1\}$, let $p_g\in\Pi$ denote the current parent program and $\mathcal L_g$ its execution feedback. Starting from $p_0=a_0$, the improvement operator $\mathcal I$ generates a set of candidate programs $\mathcal A_g$ according to
\begin{equation}
    \mathcal A_g=\mathcal I(p_g,\mathcal L_g),
    \qquad
    \mathcal L_g=\mathcal L(p_g;E),
    \label{eq:proposal}
\end{equation}
where $\mathcal L(\cdot)$ is the feedback-generation process. The feedback includes execution outcomes and may contain communication failures, action histories, and task statistics. In our implementation, $\mathcal I$ is an LLM that rewrites executable agent programs using this feedback. Each candidate is validated before performance evaluation.

The parent and valid candidates are evaluated under $E$ using a common task batch. The next parent is selected as
\begin{equation}
    p_{g+1}
    =\arg\max_{a\in\{p_g\}\cup\mathcal A_g}
    \widehat U_{E}(a),
    \label{eq:improvement}
\end{equation}
where $\widehat U_{E}(a)$ denotes the empirical utility measured under $E$, and ties favor $p_g$. Algorithm~\ref{alg:rsi} describes this process for $G$ generations with at most $C$ proposals per generation. The task batch $\mathcal B_g$ specifies the tasks used for feedback initialization or candidate evaluation at generation $g$, with $g=0,\ldots,G$. The retained program's selection logs provide the execution feedback for the next generation.

\begin{algorithm}[t]
    \caption{Wireless-in-the-loop self-improvement}
    \label{alg:rsi}
    \begin{algorithmic}[1]
        \Require Seed program $a_0$; generations $G$;
        proposals per generation $C$;
        wireless environment $E$;
        task batches $\mathcal{B}_0,\ldots,\mathcal{B}_G$
        \Ensure Final program $p_G$
        \State $p_0\gets a_0$
        \State $\mathcal{L}_0\gets\textsc{Evaluate}(p_0,\mathcal{B}_0,E)$
        \For{$g=0,\ldots,G-1$}
        \State $\mathcal{A}_g\gets\emptyset$
        \For{$c=1,\ldots,C$}
        \State $a_c\gets\textsc{Propose}(p_g,\mathcal{L}_g)$
        \State $a_c\gets\textsc{ValidateAndRepair}(a_c)$
        \If{$a_c$ is valid}
        \State $\mathcal{A}_g\gets\mathcal{A}_g\cup\{a_c\}$
        \EndIf
        \EndFor
        \State Evaluate $\{p_g\}\cup\mathcal{A}_g$
        on $\mathcal{B}_{g+1}$ under $E$
        \State $p_{g+1}\gets$ highest-scoring program
        \State \hspace{\algorithmicindent}(retain $p_g$ on ties)
        \If{$g<G-1$}
        \State $\mathcal{L}_{g+1}\gets$ selection logs of $p_{g+1}$
        \EndIf
        \EndFor
        \State \Return $p_G$
    \end{algorithmic}
\end{algorithm}

The effects of wireless conditions on this loop can be understood through the program-evolution mechanisms illustrated in Fig.~\ref{fig:mechanism}(b).

\textit{Failure exposure.}
Let $f$ denote an execution failure mode, such as an upload timeout, and let $\pi_f(p_g;E)$ denote the probability that parent $p_g$ experiences this failure in one episode under $E$. If $N$ episodes are independent, the probability that the failure occurs at least once is
\begin{equation}
    P_{\mathrm{vis}}(f)
    =1-\left(1-\pi_f(p_g;E)\right)^N.
    \label{eq:exposure}
\end{equation}
This probability characterizes whether a failure can become visible in the feedback used to generate program modifications. Under ideal communication, upload timeouts are absent, so they cannot prompt changes to the program's communication behavior.

\textit{Selection pressure.}
Candidate programs may favor different payload sizes and upload deadlines. For an upload of $b$ bits with timeout $\theta$, the probability of successful delivery under $E$ is
\begin{equation}
    P_{\mathrm{ok}}(b,\theta;E)
    ={\Pr}_{E}\left[
        \int_{t_0}^{t_0+\theta}
        r(\mathbf{x}_{t_0},u;\xi)\,{\rm d}u\ge b
        \right].
    \label{eq:pok}
\end{equation}
For a given transmission process, this probability is non-increasing in $b$ and non-decreasing in $\theta$. Let $\alpha(b)$ denote the perception accuracy associated with payload size $b$ for a specified visual content and encoding scheme. Candidate performance depends on the information conveyed by an upload, the likelihood of its completion, and the time it consumes. Evaluating programs under different wireless conditions can consequently change which communication behaviors are selected.

\textit{Program inheritance.}
Let $m$ denote a program feature, such as compact image transmission. Define $M_g(m)$ as the event that at least one candidate generated at generation $g$ exhibits $m$, and $K_g(m)$ as the event that a candidate exhibiting $m$ becomes the next parent. Given $p_g$ and the improvement environment $E$, the probability of proposing and retaining this feature is
\begin{equation}
    \begin{aligned}
        P_g(m)
        ={} & \Pr\!\left(M_g(m)\mid p_g,E\right) \\
            & \times\Pr\!\left(K_g(m)\mid
        M_g(m),p_g,E\right).
    \end{aligned}
    \label{eq:retention}
\end{equation}
The first factor captures the influence of execution feedback on candidate generation, and the second captures retention through evaluation. Since the selected program becomes the parent of the next generation, these influences also affect subsequent execution and program modifications.

The resulting search can explore new program structures. Let $\Theta$ denote the parameter space of a fixed strategy template and $h:\Theta\rightarrow\Pi$ its mapping to executable programs. Parameter optimization is confined to the programs in $h(\Theta)$, while program rewriting can modify the communication and decision-making procedures defining that template. By incorporating wireless execution feedback into program generation and selection, \sysname allows communication conditions to influence the structures explored and inherited across generations.

\subsection{Program Execution and Evaluation}

\textit{Agent execution.}
Each task-executing agent is implemented as a Python program with a \texttt{solve(env, task)} entry point and a fixed interaction interface. An auxiliary module provides reusable functions for navigation, object search, and multi-item task handling, while the agent's execution logic remains rewritable. Candidate programs can modify navigation decisions, perception choices, and communication procedures without changing the interaction interface. The task-executing program and the LLM improvement operator are separate components. Improvements are made to the program between evaluation generations and then inherited by subsequent executions.

\textit{Program generation and validation.}
The LLM improvement operator receives the parent program together with the agent interface and auxiliary functions. It also receives aggregate execution statistics and sampled execution traces. The statistics cover task outcomes, action durations, transmitted data volume, and communication failures, while the traces show successful and unsuccessful executions. The proposer observes wireless effects through these records without receiving radio maps or explicit network identities. Each proposed program must preserve the prescribed interface. Static checks verify its structure and permitted operations, followed by an execution smoke test. An invalid proposal receives one repair attempt within the same LLM interaction. Only valid programs enter candidate selection.

\textit{Wireless evaluation.}
The evaluation environment uses a shared virtual clock to account for navigation, local perception, transmission, edge inference, and waiting. During an edge query, the agent remains stationary while its achievable transmission rate changes over time. Upload completion is determined by the accumulated transmitted bits and specified timeout. Successful queries incur edge-inference and response-delivery delays. Thus, agent movement, payload choices, and query timing determine the communication outcomes recorded for program improvement.

Candidate programs are evaluated on common task batches with reproducible task, perception, and wireless randomness. Programs executing the same task share the underlying access-point placement and exogenous wireless processes, including shadowing and fading. Their actions determine where and when those processes are sampled. This evaluation design preserves the dependence between program behavior and wireless communication while supporting comparisons between candidate programs.

\section{Experimental Evaluation}
\label{sec:results}
\label{sec:method}

We evaluate \sysname to examine how wireless conditions
influence program improvement, identify the respective
roles of feedback and selection, and assess robustness
across deployment environments.

\subsection{Experimental Setup}

\begin{figure*}[t]
    \centering
    \includegraphics[width=\textwidth]{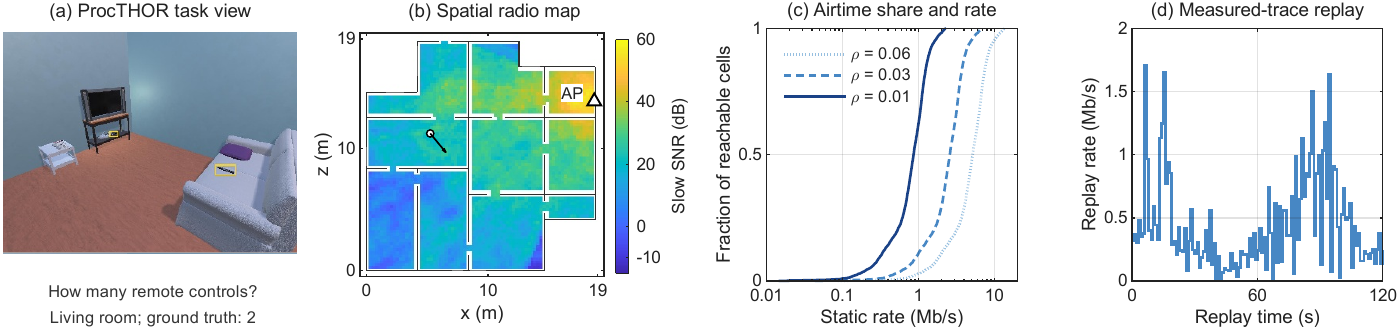}
    \caption{Embodied task environment and wireless evaluation
        conditions.
        (a) A rendered ProcTHOR scene with queried objects highlighted.
        (b) Spatial distribution of slow-varying SNR, with the
        access point and observation viewpoint marked.
        (c) Cumulative distributions of achievable link rates
        under different airtime shares.
        (d) An example of measured LTE throughput replay.}
    \label{fig:setup-env}
\end{figure*}

\textit{Embodied tasks and wireless conditions.} We instantiate \sysname in ProcTHOR indoor environments ~\cite{deitke2022procthor}, using houses with at least four rooms, divided into four disjoint sets of 20 houses for development, validation, reference evaluation, and held-out confirmation. Agents perform single-item visual question-answering tasks involving object states, counts, and spatial relationships, as well as multi-item tasks comprising \emph{inspection rounds} across three rooms and \emph{tally tasks} across multiple rooms. Each item has a 60\,s deadline. As illustrated in Fig.~\ref{fig:setup-env}, we construct spatial radio maps from house geometry on a 0.25\,m grid, incorporating path loss, shadowing, and temporal fading.
Network contention is modeled through the airtime share $\rho$ in~\eqref{eq:rate}, which scales the effective uplink transmission rate. We consider $\rho\in\{0.06,0.03,0.01\}$, corresponding to nominal, moderate, and congested conditions, respectively.
We evaluate programs under \Ideal, spatial \Radio, position-independent \Unif with matched slow-SNR statistics, and \UnifHold with link conditions retained while stationary. The \Mixed environment combines equal numbers of ideal-link and congested \Unif tasks, while \Trace uses measured LTE throughput traces scaled to corresponding analytic-channel median rates. Additional load and deadline settings are examined in sensitivity experiments.

\begin{table}[t]
    \centering
    \caption{Principal simulation parameters.}
    \label{tab:params}
    \setlength{\tabcolsep}{4pt}
    \begin{tabular}{@{}L{0.49\columnwidth}L{0.45\columnwidth}@{}}
        \toprule
        Parameter                               & Value                     \\
        \midrule
        Carrier frequency / bandwidth $W$       & 5.2\,GHz / 20\,MHz        \\
        Transmit power $P_{\mathrm{tx}}$        & 15\,dBm                   \\
        Noise figure / efficiency $\eta$        & 7\,dB / 0.6               \\
        Maximum uplink rate $r_{\max}$          & 100\,Mbit/s               \\
        Path-loss exponent $\nu$                & 3.0                       \\
        Interior / exterior wall loss           & 5 / 12\,dB                \\
        Closed-door / window loss               & 4 / 3\,dB                 \\
        Shadowing std. / decorrelation          & 4\,dB / 2\,m              \\
        Fading                                  & Rayleigh, 50\,ms blocks   \\
        Airtime share $\rho$                    &
        0.06 / 0.03 / 0.01 (main);
        0.12 (load sweep)                                                   \\
        Outages (pilot only)                    & Good / bad: 20 / 2.5\,s   \\
        \midrule
        Camera resolution                       & $1920\times1440$          \\
        Perception accuracy model               & Logistic (projected size) \\
        Max. local / edge accuracy              & 0.93 / 0.97               \\
        Object detection miss probability       & 0.2                       \\
        Median local / edge inference delay     & 0.18 / 0.35\,s            \\
        Extra edge delay per 1,000 image tokens & 0.12\,s                   \\
        Image payload size model                & JPEG-based                \\
        \midrule
        Move (0.25\,m) / rotate (90$^\circ$)    & 0.5 / 0.3\,s              \\
        Default upload timeout                  & 2\,s                      \\
        Task deadline                           & 60\,s (main);
        20 / 40\,s (sensitivity)                                            \\
        \bottomrule
    \end{tabular}
\end{table}

\textit{Perception, baselines, and selection.}
To support large-scale evaluation, we use analytic perception models that relate accuracy to target visibility and projected image size, with separate models for inference delays and image payload sizes. We further examine these assumptions using a real VLM on rendered task images. Table~\ref{tab:params} summarizes the simulation parameters. We construct two pools of 16 hand-designed programs for single-item and multi-item tasks, respectively, comprising six base strategies and variants in navigation, perception, communication, and retry decisions. Sharing a common visual search procedure, the base strategies differ in their use of local and edge perception. Full-frame uploads complete camera images, while always-crop transmits target-centered crops based on simulated detections. Deferred-upload also uses cropped images but adjusts transmission timing and location based on observed signal strength. Local-first invokes edge perception with cropped images only when local confidence is insufficient. These four strategies fall back to local answers upon edge-query failures, while move-closer and local-only require no uploads. Fixed-program experiments vary network load, access-point placement, outages, task deadlines, and perception assumptions. For fixed-pool selection, each candidate is evaluated on 200 validation tasks per network condition, with selection budgets of 8--64 episodes per candidate. All 16 candidates are independently evaluated on 400 reference tasks under each deployment condition to determine the empirical fixed-pool oracle and selection regret.

\textit{Self-improvement and parameter optimization.}
Each self-improvement lineage starts from the same multi-item full-frame program and evolves over four generations, with two LLM proposals per generation. The parent and valid children are evaluated on a fresh batch of 40 tasks using shared episode seeds, retaining the parent on ties. Unlike fixed-pool selection, self-improvement can generate programs beyond predefined execution strategies. We use \texttt{gpt-6.1-terra} as the primary proposer, with eight paired lineages under six improvement environments: \Ideal, \Radio at $\rho=0.06$, $0.03$, and $0.01$, \Unif at $\rho=0.01$, and \Mixed. Each improvement environment governs both execution feedback and candidate selection. We also evaluate \texttt{gpt-6.1-sol} with six lineages each under ideal-link and congested \Radio improvement. All intermediate parents and final programs are independently audited on 400 reference tasks under the ideal link and three wireless loads. To examine whether these gains can be achieved by tuning predefined strategies, we construct an eight-parameter template covering perception thresholds, payload choices, timeouts, retries, movement, and link-dependent fallback, yielding 14,400 configurations. We compare random parameter mutation under the same evaluation budget and schedule as WiSI with successive halving over 65 initial configurations using approximately eleven times the budget.

\textit{Mechanism analysis and validation.} To distinguish the effects of feedback and selection, we independently control the feedback environment $E_L$, which supplies execution experience for program generation, and the selection environment $E_S$, which governs candidate evaluation. We conduct a $2\times2$ factorial experiment with $E_L,E_S\in\{\Ideal,\Radio\}$, using congested \Radio and eight paired lineages per configuration. All configurations share the same proposal and selection budgets, with additional feedback executions when $E_L\neq E_S$. Initial feedback is collected under $E_L$. For subsequent generations, the retained program's selection logs are reused when $E_L=E_S$. Otherwise, the retained program is re-evaluated under $E_L$ on the same task batch used for selection. We compare program-level improvement with parameter optimization over a predefined strategy template containing 14,400 configurations, using both matched and enlarged evaluation budgets. We further examine robustness to inaccurate run-time link observations and validate the findings using measured LTE traces, real-VLM perception, and 20 held-out houses. All results are reported with 95\% confidence intervals, using house-cluster bootstrap for fixed-program evaluations and paired lineage bootstrap with 10,000 resamples for program-improvement comparisons. Held-out evaluations additionally account for house-level variability.
For held-out validation, we pre-specify eight decision criteria covering wireless-improvement gains under congested, nominal, and moderate loads; improvements over hand-designed strategies; replication with a second LLM proposer; transfer to ideal-link deployment; mixed-network performance; and simplified channel modeling. The mixed-network criterion involves two comparisons under ideal and congested deployment. All decision margins are fixed before held-out evaluation.

\subsection{Experimental Results}

\subsubsection{Effectiveness of Wireless-in-the-Loop Improvement}
\label{sec:res-regimes}
\label{sec:res-trajectory}

\begin{table}[t]
    \centering
    \caption{Task success (\%) of final programs under different deployment conditions, averaged over eight improvement lineages. Worst denotes the mean, across eight lineages, of each
        program's minimum success rate over the four deployment
        conditions. Link indicates the number of programs accessing run-time link observations.}
    \label{tab:main}
    \setlength{\tabcolsep}{2.4pt}
    \begin{tabular}{@{}lcccccc@{}}
        \toprule
        Improved under    & ideal         & 0.06          & 0.03          & 0.01          & worst         & link \\
        \midrule
        \Ideal            & 82.2          & 78.2          & 69.6          & 40.0          & 40.0          & 0/8  \\
        \Radio, 0.06      & \textbf{84.0} & \textbf{82.8} & \textbf{81.8} & 75.1          & 75.1          & 3/8  \\
        \Radio, 0.03      & 83.6          & 82.4          & 81.6          & 75.2          & 75.2          & 5/8  \\
        \Radio, 0.01      & 79.0          & 78.5          & 78.2          & 77.2          & 77.0          & 3/8  \\
        \Unif, 0.01       & 79.0          & 78.5          & 77.9          & 76.1          & 76.1          & 2/8  \\
        \Mixed            & 83.0          & 80.1          & 79.8          & \textbf{77.5} & \textbf{77.5} & 8/8  \\
        \midrule
        Seed (full frame) & 73.8          & 69.8          & 58.8          & 24.5          & 24.5          & --   \\
        Best hand-written & 73.8          & 72.2          & 70.5          & 53.0          & 53.0          & --   \\
        \bottomrule
    \end{tabular}
\end{table}

\begin{figure*}[t]
    \centering
    \includegraphics[width=\textwidth]{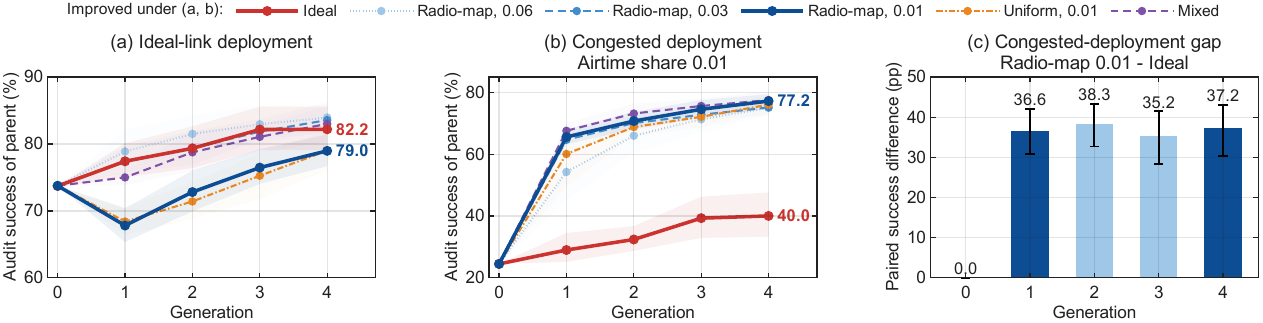}
    \caption{Evolution of program performance across improvement generations.
        (a), (b) Parent-program success under ideal-link and congested deployment, respectively, averaged over eight lineages per evaluation environment.
        (c) Paired congested-deployment success differences between programs improved under congested \Radio and \Ideal.
        Shaded regions and error bars indicate 95\% lineage-bootstrap confidence intervals, with paired resampling used in (c).}
    \label{fig:generations}
\end{figure*}

We first examine whether incorporating wireless conditions into the improvement process changes the programs ultimately obtained. Table~\ref{tab:main} reveals a substantial dependence on the improvement environment. Programs evolved under \Ideal achieve only 40.0\% task success when deployed under congestion, while those evolved under congested \Radio achieve 77.2\%, an advantage of 37.2 percentage points ($[30.3,43.0]$). The latter also outperform the best hand-designed program by 24.2 points ($[21.9,26.4]$). Fig.~\ref{fig:generations} shows that this divergence emerges early. The congested-deployment advantage reaches 36.6 points ($[30.9,42.1]$) after the first improvement generation and persists through the fourth. Independent audits also show that 94\% of accepted child programs improve reference-task performance. The improvement environment thus affects the trajectory of program evolution from its initial iterations, with consequences for the behavior retained in subsequent generations.

\begin{figure*}[t]
    \centering
    \includegraphics[width=\textwidth]{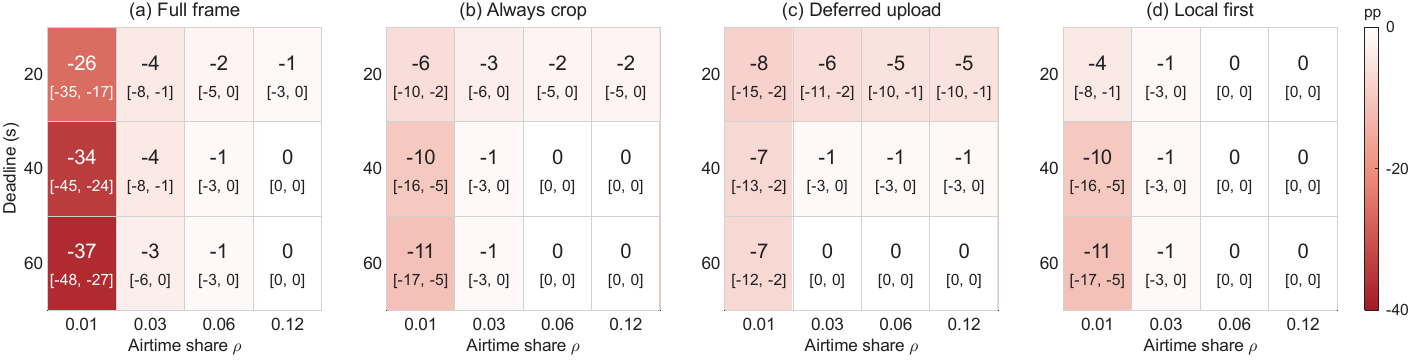}
    \caption{Impact of wireless conditions on fixed-program execution. Each heatmap reports the difference in task success between \Radio and \Ideal across task deadlines and airtime shares. Cell annotations give the point estimates in percentage points and their 95\% house-bootstrap confidence intervals.}
    \label{fig:load}
\end{figure*}
To understand the role of wireless evaluation before program rewriting, we examine four fixed strategies under different network loads and task deadlines. As shown in Fig.~\ref{fig:load}, the full-frame strategy exhibits little degradation under nominal load, where navigation dominates task duration. Under severe congestion ($\rho=0.01$), however, its success drops by 26--37 percentage points, compared with 4--11 points for the three compact-upload strategies. Their residual losses arise primarily when missed detections trigger fallback to full-frame uploads, highlighting the importance of payload size in determining wireless sensitivity. In addition, RSSI-based deferred upload offers limited benefits against contention and incurs additional delay under tight deadlines. These results demonstrate that wireless conditions can substantially change the relative performance of existing execution strategies, making realistic evaluation important for subsequent candidate selection.

\begin{figure*}[t]
    \centering
    \includegraphics[width=\textwidth]{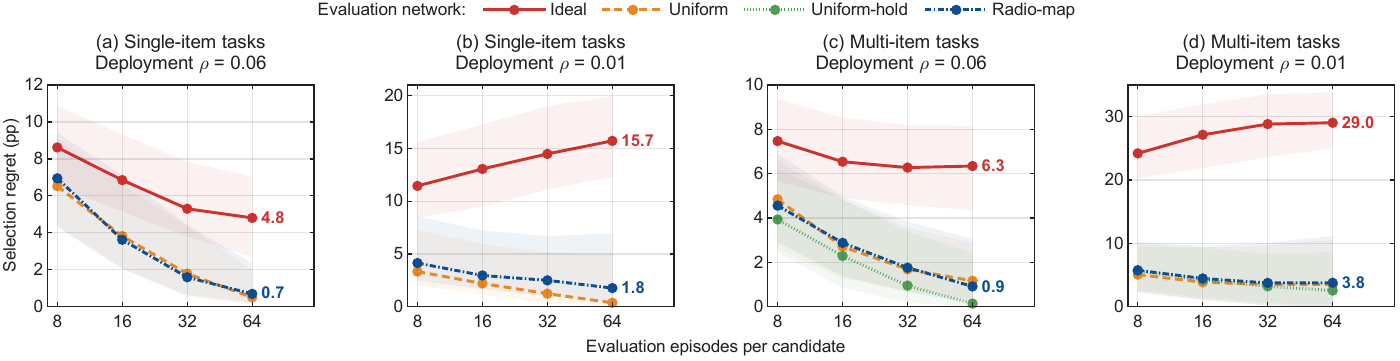}
    \caption{Deployment selection regret under different evaluation networks and budgets. Results cover single-item and multi-item tasks under nominal and congested deployment. Shaded regions denote 95\% joint house-bootstrap confidence intervals.}
    \label{fig:selection}
\end{figure*}
Fig.~\ref{fig:selection} confirms this selection mismatch. Increasing the evaluation budget reduces uncertainty when the evaluation and deployment conditions match, yet even 64 episodes per candidate leave 29.0 percentage points of selection regret for multi-item tasks evaluated under \Ideal and deployed under congestion, compared with 3.8 points under matched \Radio evaluation. More accurate estimates of ideal-link performance cannot correct a mismatch in which programs the environment favors. The fixed-program results establish why the wireless evaluation environment matters; the self-improvement results show that its influence extends to the programs generated over successive iterations.

\subsubsection{Mechanisms of Wireless-Dependent Program Evolution}
\label{sec:res-channels}
\label{sec:res-mechanism}

\begin{figure*}[t]
    \centering
    \includegraphics[width=\textwidth]{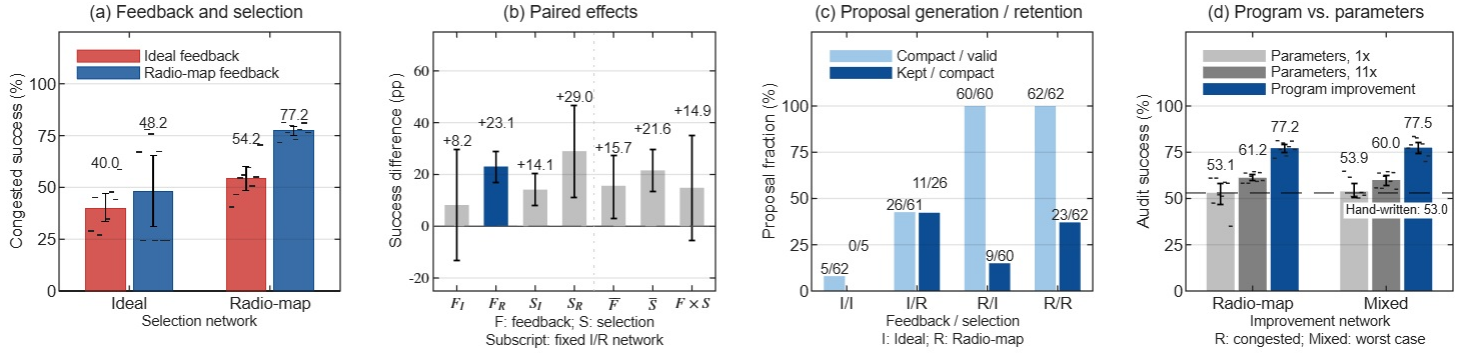}
    \caption{Wireless feedback and selection shape program evolution.
        (a) Final task success under independently controlled feedback and selection environments.
        (b) Conditional feedback and selection effects, their averages, and interaction, with 95\% paired lineage-bootstrap confidence intervals.
        (c) Code-audited compact-program proposal and retention frequencies across the four experimental conditions.
        (d) Comparison of program-level improvement and parameter optimization under matched and enlarged evaluation budgets.}
    \label{fig:channels}
\end{figure*}

\begin{figure*}[t]
    \centering
    \includegraphics[width=0.8\textwidth]{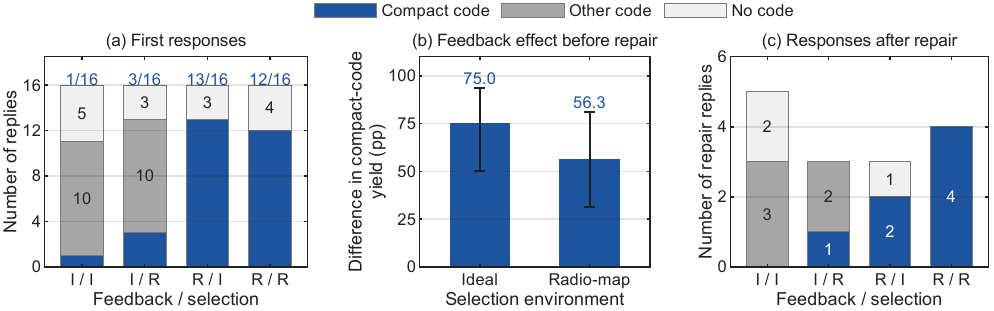}
    \caption{First-generation proposal responses under different feedback and selection environments.
        (a) Initial LLM responses classified as compact code, other code, or no code.
        (b) Feedback effects on compact-code generation per proposal slot, with 95\% paired lineage-bootstrap confidence intervals.
        (c) Outcomes of repair attempts for responses without code.
        Results are from a post-hoc audit of eight paired lineages per condition.}
    \label{fig:first-generation}
\end{figure*}

We next separate the effects of execution feedback and candidate selection through a $2\times2$ factorial experiment with $E_L,E_S\in\{\Ideal,\Radio\}$, using congested \Radio and eight paired lineages per condition. All conditions share the same proposal and candidate-selection budgets, while off-diagonal conditions require additional feedback evaluations under $E_L$. As shown in Fig.~\ref{fig:channels}(a), using realistic wireless conditions for either feedback or selection improves final performance, with the largest gain achieved when both operate under congestion. Importantly, even when selection already uses congested \Radio, changing feedback from \Ideal to \Radio improves final task success by 23.1 percentage points ($[16.8,28.7]$). Fig.~\ref{fig:channels}(b) further reports the average effects and their interaction, whose confidence interval includes zero. These results demonstrate that wireless execution feedback influences the improvement trajectory beyond its effect on candidate selection.

The program audit clarifies where this contribution arises. Under realistic feedback, compact communication structures appear in nearly every valid proposal regardless of the selection environment, compared with only 5 of 62 when both environments are ideal (Fig.~\ref{fig:channels}(c)). When feedback is held fixed to \Radio, realistic selection retains 23 of 62 compact candidates, compared with 9 of 60 under ideal selection. To isolate candidate generation from subsequent selection and inheritance, we further audit the first LLM responses in generation one. All conditions share the same seed program and non-feedback prompt content. With selection fixed to congested \Radio, compact code appears in 3 of 13 code-containing first responses under ideal feedback and 12 of 12 under congested feedback. Across all 16 proposal slots per condition, this is a 56.3-percentage-point increase ($[31.3,81.3]$; Fig.~\ref{fig:first-generation}). The difference is already present before validation, repair, selection, or changes to the parent program. Wireless execution feedback therefore influences the structures entering the program search, while selection governs their retention across generations.

\subsubsection{Communication Efficiency of Evolved Programs}
\label{sec:res-structures}

\begin{figure*}[t]
    \centering
    \includegraphics[width=\textwidth]{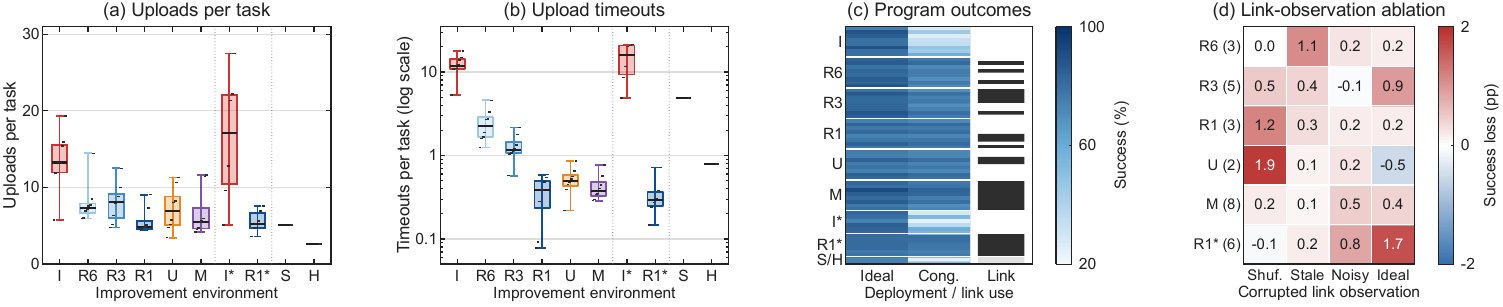}
    \caption{Communication behavior and robustness of evolved programs.
        (a) Distribution of uploads per task under congestion.
        (b) Distribution of upload timeouts per task on a logarithmic scale.
        (c) Ideal-link and congested-deployment success of individual programs, with link-observation usage indicated separately.
        (d) Mean task-success loss under corrupted run-time link observations.
        I denotes \Ideal; R6, R3, and R1 denote \Radio at airtime shares 0.06, 0.03, and 0.01; U denotes \Unif; M denotes \Mixed. Asterisks distinguish the second proposer, while S and H indicate the seed and hand-designed reference.}
    \label{fig:economy}
\end{figure*}

We next investigate whether the communication efficiency of evolved programs stems from better parameter choices within predefined strategies or from more effective coordination of perception, transmission, and fallback decisions. Fig.~\ref{fig:channels}(d) compares WiSI with parameter optimization over a predefined strategy template. Under congested \Radio, parameter optimization achieves 53.1\% task success with a matched evaluation budget and 61.2\% with approximately eleven times that budget, compared with 77.2\% for program-level improvement. Similar advantages are observed under \Mixed, where program-level improvement achieves 77.5\% worst-case success, compared with 53.9\% and 60.0\% for the two parameter-optimization baselines. While parameter optimization adjusts predefined execution strategies, program-level improvement can reorganize how perception, communication, and fallback decisions are made. These results suggest that structural flexibility provides benefits beyond parameter tuning within the evaluated template and budgets.

To understand the communication behaviors underlying these gains, Fig.~\ref{fig:economy}(a) and (b) examine upload frequency and transmission failures. Under the same congested deployment, programs improved over ideal links typically attempt 13--16 uploads and encounter roughly 12--15 timeouts per task. In contrast, wireless-improved programs generally require only 5--9 uploads, with timeout counts falling below 0.6 under congested \Radio, \Unif, and \Mixed improvement. Together with the compact communication structures identified in Fig.~\ref{fig:channels}(c), these results show how wireless-aware improvement reshapes the communication behavior of evolved programs. Importantly, these communication-efficient behaviors do not necessarily require continuous run-time channel sensing. Fig.~\ref{fig:economy}(c) reveals strong congested-deployment performance even among programs that do not access link observations. For the 27 programs that use such observations, the tested observation corruptions cause at most 1.9 percentage points of mean success loss (Fig.~\ref{fig:economy}(d)). These findings indicate that wireless-aware improvement can encode communication-efficient decisions directly into evolved programs, with limited dependence on precise run-time link observations.

\subsubsection{Robustness Across Wireless Conditions}
\label{sec:res-loadmix}
\label{sec:res-fidelity}

\begin{figure}[t]
    \centering
    \includegraphics[width=0.9\columnwidth]{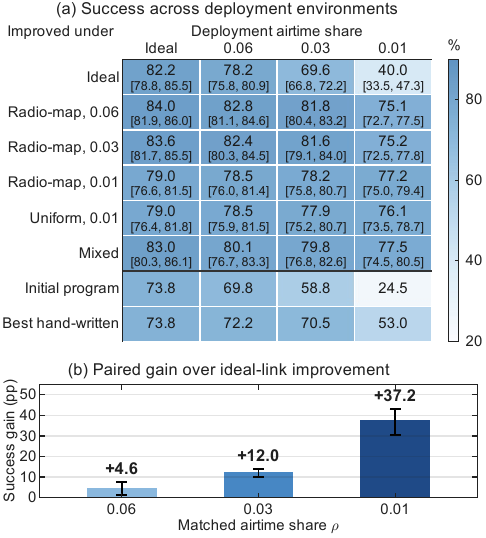}
    \caption{Cross-load robustness under different wireless improvement environments.
        (a) Task success across deployment conditions, with 95\% lineage-bootstrap confidence intervals for evolved programs.
        (b) Paired success gains of matched \Radio improvement over \Ideal improvement at different airtime shares, with 95\% paired lineage-bootstrap confidence intervals.}
    \label{fig:loadmix}
\end{figure}

The improvement environment also determines how well an evolved program transfers across deployment loads. Fig.~\ref{fig:loadmix} and Table~\ref{tab:main} show that programs improved under nominal or moderate contention retain strong ideal-link performance while achieving substantial gains under severe congestion. The benefit of matched wireless improvement grows with deployment contention, from 4.6 percentage points at $\rho=0.06$ to 12.0 and 37.2 points at $\rho=0.03$ and $0.01$, respectively. Notably, improvement under moderate wireless contention already produces programs that transfer effectively to more congested environments, without requiring training under the most severe network conditions. Conversely, improving exclusively under severe congestion sacrifices some performance under lighter loads. The \Mixed environment provides a more balanced outcome, achieving the highest observed mean worst-case success across the four deployment conditions. These findings highlight the importance of selecting improvement environments that expose communication constraints while preserving effective performance across heterogeneous deployment loads.

We further investigate whether effective program improvement requires detailed spatial channel modeling. The \Unif environment preserves the marginal slow-SNR distribution of the radio map while removing spatial dependence. Programs improved under \Unif achieve similar congested-deployment performance to those improved under spatial \Radio, with a paired difference of $-1.2$ percentage points ($[-4.6,2.0]$). Interestingly, similar final performance does not imply identical communication dynamics. For example, retransmitting at the same location after a timeout succeeds considerably more often under \Unif than under \Radio or the temporally persistent \UnifHold model. Despite these differences in local communication behavior, the resulting programs achieve comparable task performance under congestion. This suggests that detailed spatial radio maps are not always necessary for effective program improvement. In the evaluated tasks, preserving relevant link-quality statistics can provide a lower-complexity alternative for generating useful wireless execution feedback.

\subsubsection{Independent Validation and Generalization}
\label{sec:res-heldout}

\begin{figure*}[t]
    \centering
    \includegraphics[width=\textwidth]{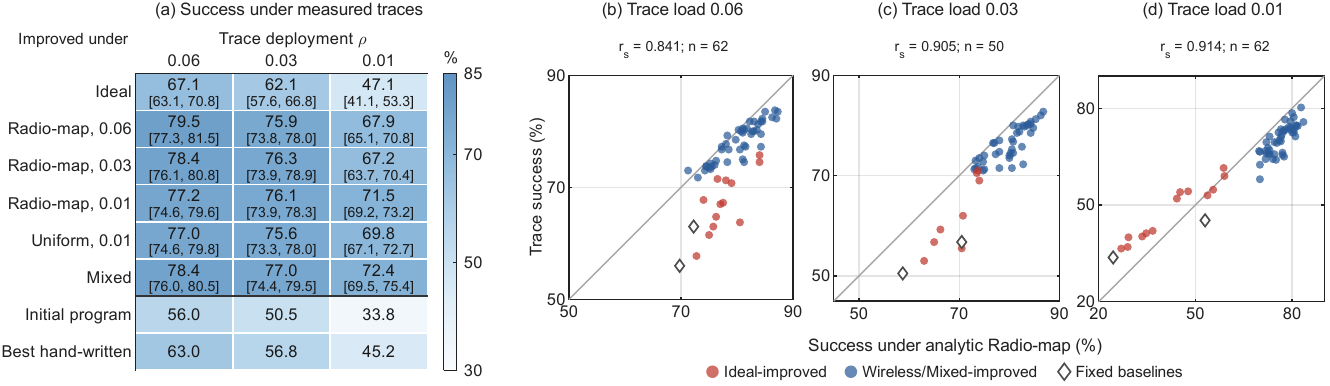}
    \caption{Validation under measured LTE throughput traces.
        (a) Task success across trace-driven deployment conditions, with 95\% lineage-bootstrap confidence intervals for improved programs.
        (b)--(d) Program-level comparisons between analytic and measured channels at three corresponding network loads. Each point represents one program; colors distinguish improvement environments, and diamonds indicate fixed baselines. The diagonal denotes equal performance, and $r_s$ is Spearman rank correlation.}
    \label{fig:trace}
\end{figure*}

\begin{figure*}[t]
    \centering
    \includegraphics[width=\textwidth]{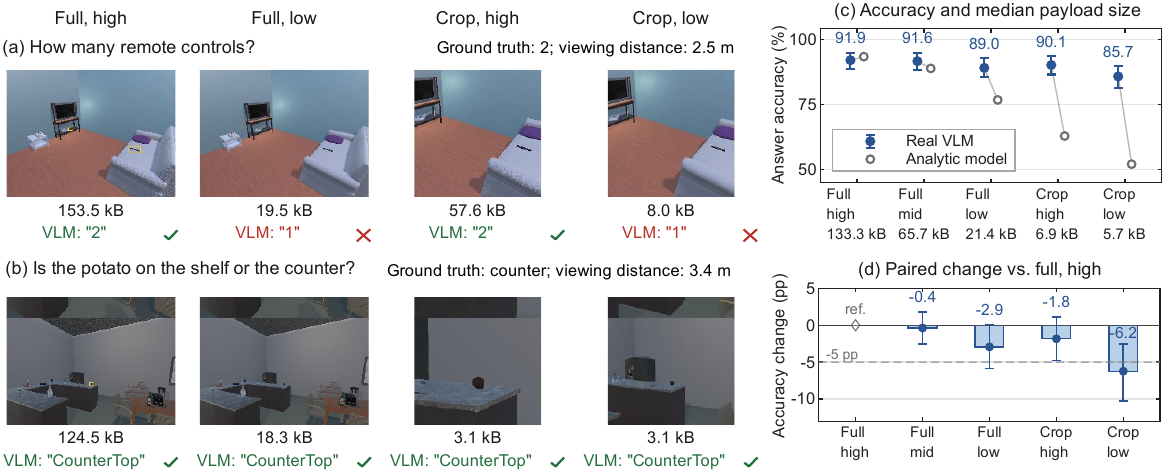}
    \caption{Real-VLM perception performance under different communication payloads.
        (a), (b) Representative counting and spatial-relation tasks with different image payloads and model responses.
        (c) Real-VLM accuracy and analytic-model predictions across five payload configurations, with median transmitted sizes and 95\% item-bootstrap confidence intervals.
        (d) Paired accuracy changes relative to high-resolution full frames, with 95\% paired item-bootstrap confidence intervals. The dashed line indicates the five-percentage-point accuracy-loss threshold.}
    \label{fig:vlm}
\end{figure*}

\begin{figure}[t]
    \centering
    \includegraphics[width=\columnwidth]{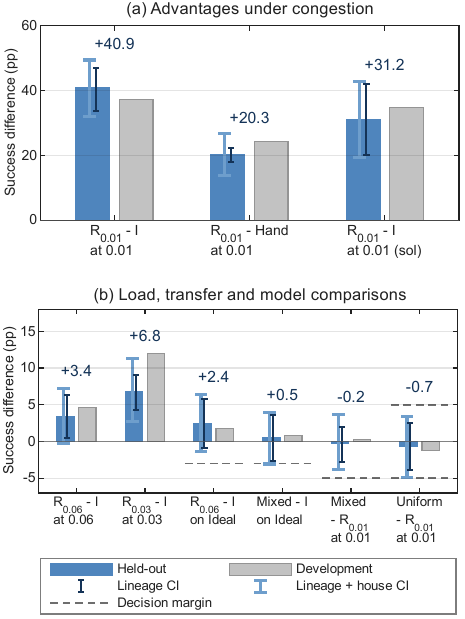}
    \caption{Nine held-out performance contrasts evaluating eight pre-specified decision criteria on 20 unseen houses.
        (a) Performance advantages under congested deployment.
        (b) Load, transfer, and channel-model comparisons.
        Blue and gray bars denote held-out and development estimates, respectively. Dark and light error bars show 95\% lineage-bootstrap and joint lineage-and-house bootstrap confidence intervals for held-out estimates. Dashed markers indicate pre-specified decision margins.}
    \label{fig:heldout}
\end{figure}

The preceding experiments show that wireless execution conditions shape program evolution, leading to more communication-efficient behaviors and improved performance across deployment loads. We now investigate whether these findings extend beyond the analytic wireless and perception models used in the main experiments. To this end, we conduct three independent validations using measured LTE throughput traces, real-VLM perception, and previously unseen indoor environments.

We first examine whether the findings transfer from analytic wireless channels to measured communication conditions. We replay 31 pedestrian LTE throughput traces from a public dataset~\cite{raca2018lte} using trace-driven emulation~\cite{netravali2015mahimahi}, with disjoint improvement and audit pools and rates scaled to the corresponding analytic-channel medians. Under congested trace-driven deployment, programs improved under \Radio outperform ideal-link-improved programs by 24.4 percentage points ($[18.5,30.0]$; Fig.~\ref{fig:trace}(a)), while programs improved directly on traces achieve comparable performance. Despite differences in absolute success rates and deep-fade behavior, analytic and trace-driven program rankings remain strongly correlated across deployment loads, with Spearman coefficients between 0.841 and 0.914 (Fig.~\ref{fig:trace}(b)--(d)). These results suggest that analytic wireless models can provide useful guidance for program improvement by preserving the relative merits of candidate programs, even without reproducing all characteristics of measured communication conditions.

We then examine whether compact communication remains effective with a real vision-language model (VLM). Across 273 rendered AI2-THOR task items, high-resolution full frames achieve 91.9\% perception accuracy at a median payload of 133.3\,kB, while high-resolution crops retain 90.1\% accuracy at only 6.9\,kB (Fig.~\ref{fig:vlm}). Notably, high-resolution crops also achieve higher observed accuracy than low-resolution full frames (89.0\% at 21.4\,kB), suggesting that preserving task-relevant visual regions can be more effective than uniformly reducing image resolution. These results demonstrate a favorable tradeoff between communication overhead and perception quality, while also revealing the limitations of excessive compression. Since the analytic perception model does not perfectly match these measurements, we recalibrate it using real-VLM results and re-audit the existing programs. The main congested-deployment advantage over ideal-link improvement persists at 40.6 percentage points ($[32.5,46.9]$), indicating that the benefits of wireless-aware program evolution extend beyond the original perception approximation.

Finally, we examine whether the evolved programs generalize to previously unseen physical environments. We independently audit 62 programs on 20 held-out ProcTHOR houses that were excluded from development and evaluation. Fig.~\ref{fig:heldout} reports nine performance comparisons covering eight pre-specified decision criteria, with two comparisons jointly assessing \Mixed under ideal and congested deployment. Under congestion, programs improved under \Radio retain a 40.9-percentage-point advantage over ideal-link-improved programs and outperform the best hand-designed baseline by 20.3 points (Fig.~\ref{fig:heldout}(a)). The main advantage also persists with a second LLM proposer. All eight criteria are satisfied under lineage-bootstrap analysis, while joint resampling of lineages and houses supports the main congestion-related gains and reveals greater uncertainty in smaller nominal-load effects (Fig.~\ref{fig:heldout}(b)). Together, these independent evaluations support the principal findings across measured communication conditions, real-VLM perception, and unseen indoor environments.

\section{Conclusion}\label{sec:conclusion}

This paper investigated how wireless communication shapes the embodied AI agent's self-improvement. We developed \sysname, a wireless-in-the-loop framework that incorporates communication conditions into iterative program improvement and distinguishes their roles in execution feedback and candidate selection.
Our experiments show that wireless conditions shape both candidate generation and program retention. Realistic wireless feedback improves congested-deployment success by 23.1 percentage points even when candidate selection is fixed. Overall, wireless-in-the-loop improvement achieves 77.2\% success under congestion, compared with 40.0\% for ideal-link improvement, producing communication-efficient programs that outperform parameter optimization within predefined strategy templates. Moderate-load and mixed-network improvement further support effective transfer across deployment conditions, while simplified wireless models yield comparable results in the studied tasks. These findings are further supported by measured LTE traces, real-VLM perception, and held-out environments.
Future work will extend wireless-in-the-loop improvement to physical robotic platforms and more complex communication settings, including mobility during transmission and adaptive improvement operators.
These findings highlight a fundamental role of wireless communication in self-improving AI systems, where communication conditions shape not only how agents execute tasks but also how their programs evolve.

\bibliographystyle{IEEEtran}
\bibliography{refs}

\end{document}